\documentclass[5p]{elsarticle}

\usepackage{ecrc}

\volume{00}

\firstpage{1}

\journalname{Astronomy and Computing}

\runauth{Farid et al.}

\jid{procs}

\jnltitlelogo{Astronomy and Computing}

\usepackage{graphicx}	
\usepackage{amsmath}	
\usepackage{amssymb}	
\usepackage{xcolor}
\PassOptionsToPackage{hyphens}{url}\usepackage{hyperref}
\usepackage{cleveref}
\usepackage{ulem}
\usepackage{multirow} 

\newcommand{\ave}[1]{\left\langle #1\right\rangle}

\newcommand{\Msunh}{M_{\odot}/h}
\newcommand{\vlos}{v_{\rm los}}
\def\r200m{r_{\rm 200m}}

\usepackage[figuresright]{rotating}

\begin{document}

\begin{frontmatter}




\title{\textsc{C$^2$-GaMe}: Classification of Cluster Galaxy Membership with Machine Learning}


\author[1]{Daniel Farid}
\author[2]{Han Aung}
\author[2]{Daisuke Nagai}
\author[3]{Arya Farahi}
\author[4]{Eduardo Rozo}

\affiliation[1]{organization={Program in Applied Mathematics, Yale University},
postcode={CT 06511},
city={New Haven},
country={USA}}

\affiliation[2]{organization={Department of Physics, Yale University},
postcode={CT 06520},
city={New Haven},
country={USA}}


\affiliation[3]{organization={Department of Statistics and Data Science, The University of Texas},
postcode={TX 78712},
city={Austin},
country={USA}}

\affiliation[4]{organization={Department of Physics, University of Arizona},
postcode={AZ 85721},
city={Tucson},
country={USA}}

\begin{abstract}
We present \textsc{C}lassification of \textsc{C}luster \textsc{Ga}laxy \textsc{Me}mbers (\textsc{C$^2$-GaMe}), a classification algorithm based on a suite of machine learning models that differentiates galaxies into orbiting, infalling, and background (interloper) populations, using phase space information as input. We train and test \textsc{C$^2$-GaMe} with the galaxies from UniverseMachine mock catalog based on Multi-Dark Planck 2 N-body simulations. We show that probabilistic classification is superior to deterministic classification in estimating the physical properties of clusters, including density profiles and velocity dispersion. We propose a set of estimators to get an unbiased estimation of cluster properties.
We demonstrate that \textsc{C$^2$-GaMe} can recover the distribution of orbiting and infalling galaxies' position and velocity distribution with $<1\%$ statistical error when using probabilistic predictions in the presence of interlopers in the projected phase space. Additionally, we demonstrate the robustness of trained models by applying them to a different simulation. Finally, adding a specific star formation rate and the ratio of the galaxy's halo mass to the cluster's halo mass as additional features improves the classification performance. We discuss potential applications of this technique to enhance cluster cosmology and galaxy quenching. 
\end{abstract}

\end{frontmatter}


\section{Introduction}

Galaxy clusters reside in the most massive gravitationally bound dark matter halos. They are unique laboratories for measuring the gravitational interactions in the universe that directly influence the collapse and growth of these large-scale structures \citep{kravtsov_borgani}. Upcoming spectroscopic surveys, such as DESI, promise to provide unprecedented data, enabling spectroscopy of hundreds of thousands of galaxies \citep{Desi}. Spectroscopic measurements of galaxies around galaxy clusters allow us to estimate the dynamics of these galaxies. Dynamics of cluster galaxies, such as the velocity dispersion inside the clusters \citep{evrard08,bocquet15} or the infall velocity into the clusters \citep{hamabata19}, are potentially powerful mass proxies. They are also unique probes of the large-scale inflow to probe modified gravity \citep{lam12,zu2014}. However, using these dynamical measurements for cosmology requires a detailed understanding and control of the associated systematic uncertainties. 

The phase space structure of the halos consists of kinematically distinct populations of infalling and orbiting galaxies \citep{Aung2021}. However, the spatial distinction to include all orbiting galaxies extends much further than the traditional radii definition as the backsplash galaxies exist beyond virial radii \citep{Balogh2000,Mamon2004,Gill2005,ludlow2009,wang2009}, and the galaxies inside the cluster radii can also be on their first infall. Observationally, these two populations are mixed along with interlopers, which are in front or behind the halo of interest but appear closer because of the projection and blend into the cluster we want to study \citep[e.g.,][]{vdb04,biviano06}. Any contaminating interlopers can significantly bias the dynamical mass estimate by 10-15$\%$ \citep{Wojtak18}. Several methods for removing interlopers by treating individual clusters separately \citep{Wojtak07} or from a stacked sample of samples \citep{rozo2015} exist in the literature. These methods cannot remove all interlopers while retaining all the member galaxies, but can recover dynamical mass in good agreement with other probes with correct treatment of interlopers and galaxy velocity bias \citep{Farahi16, farahi2018xxl}. 

In addition, measuring the radial distributions of the orbiting galaxies allows us to infer the edge radius, which includes all orbiting galaxies. The edge radius can be a standard ruler for measuring Hubble constant \citep{wagoner2021}. A proxy for mass accretion rate similar to the splashback radius \citep{Diemer2014, more_etal2015, Aung2021} as mass accretion rate can predict secondary halo properties such as concentration \citep{Wechsler2002,ludlow2013}, star formation history \citep{Diemer2013,Wetzel2015}, and non-thermal pressure \citep{Shi2014,Green2020}.
 
Machine learning (ML) methods have become an increasingly powerful tool in the era of data-intensive cosmology and astrophysics \citep{Baron19,mlcosmo_Ntampaka19}. They are especially effective in classification problems, such as classifying different types of galaxies using star formation and morphological properties \citep{class_Baldeschi20}, type Ia supernova \citep{class_lochner16}, sources in reionization era \citep{class_Hassan19}, and planets \citep{class_pham21}. The ML algorithms employed in these studies achieve accuracy rates exceeding 90\%, highlighting their robust performance. Moreover, they have exhibited comparable or superior performance when compared to traditional methods, demonstrating their capability to effectively handle complex astrophysical datasets. Additionally, ML methods offer the advantage of enhanced scalability, allowing for efficient processing of large-scale datasets commonly encountered in cosmology and astrophysics research. More closely, ML algorithms allow for the classification of galaxy membership within galaxy clusters through photometric or spectroscopic information \citep{Lopes2020,Angora2020,Hashimoto2022}.
\citet{delosRios2021} used several classifiers to classify galaxies into five classes using 2D phase space information. They found that random forest recreates the distributions better, while k-nearest neighbor (KNN) produces better precision and recall.  

This work uses machine learning models to classify galaxy populations in the UniverseMachine mock galaxy catalog of N-body simulation MDPL2. We compare probabilistic membership assignments with deterministic ones and introduce a new set of features to improve the classification performance and robustness to enable scientific applications. Our findings include (a)  probabilistic weighting, which utilizes uncertainty in prediction instead of summarizing the classification as a single class, can improve the estimation of cluster physical properties, (b) identifying additional features with the potential to improve the classification of orbiting and infalling galaxies, (c) robustness of the galaxy classification method is tested on the full Illustris-TNG hydrodynamical cosmological simulations, and (d) demonstration that our probabilistic ML approach can reproduce the radial distributions with $<3\%$ uncertainty for measuring edge radius and predict velocity dispersion with $<1\%$ bias required for dynamical mass estimation.

In this work, we present \textsc{C}lassification of \textsc{C}luster \textsc{Ga}laxy \textsc{Me}mbers (\textsc{C$^2$-GaMe}) for classifying galaxies into orbiting, infalling, and background (interloper) populations. We describe the simulation data sets used to train and test the ML algorithms in Section~\ref{sec:sim}. We detail the property estimation based on the classification of the ML algorithm in Section~\ref{sec:ML}. We describe the evaluation methods used to determine the performance of our model in Section~\ref{sec:evaluation_methods}. We report results in Section~\ref{sec:results}. We provide implications and potential applications in Section~\ref{sec:application} and our main conclusion in Section~\ref{sec:conc}. We publicly release \textsc{C$^2$-GaMe}, a Python package hosted on a GitHub repository\footnote{\url{https://github.com/DannyFarid/C2-GaMe-Classification-of-Cluster-Galaxy-Membership}}.

\section{Simulation Data}
\label{sec:sim}

\subsection{Cluster and Galaxy Catalogs}

In this work, we analyze the outputs of the MultiDark Planck 2 (MDPL2) dark-matter-only N-body simulation, performed using L-GADGET-2 code (a version of the publicly available cosmological code GADGET-2 \citep{Springel2005}). It contains $3840^3$ total dark matter particles, where the mass of one individual particle is $1.51\times 10^{9} \mathrm{M}_{\odot} / \mathrm{h}$. The simulation uses Planck13 cosmology, where $h  = 0.6777$,  $\Omega_{\Lambda} = 0.692885$, $\Omega_m = 0.307115$, $\Omega_{b} = 0.048206$, $n = 0.96$, and $\sigma_8 = 0.8228$ \citep{planck13}. The halos and subhalos hosting the galaxies are identified using the Rockstar halo catalog \citep{Behroozi2013_rs}, and the merger tree is built using the Consistent Tree algorithm \citep{Behroozi2013}. We select central halos with $M_{200m}>10^{14} M_{\odot} / h$ which are not within other larger halos (Rockstar primary host), and satellite galaxies with a peak mass of $M_{p}>3 \times 10^{11} M_{\odot} / h$ at redshift of $z=0.5$. This results in $\sim 17000$ clusters with at least 20 galaxies within $r_{\rm 200m}$ of every central halo. 

The mock galaxy catalog is constructed using the UniverseMachine \citep{Behroozi19}, in which galaxies are pasted onto halos and subhalos such that the low-order statistical properties of the resulting galaxy properties match observational data across cosmic times.
The in-situ star formation rate is parameterized as a function of halo mass, halo assembly history, and redshift. We compute the stellar mass of the halo by integrating the star formation rates over the merger history of the halo. The mock catalog reproduces the following observational data across a broad range of redshifts ($0<z<10$): (i) stellar mass functions; (ii) cosmic star-formation rates and specific star-formation rates; (iii) quenched fractions\footnote{fractions of galaxies that are not star forming, usually with specific star formation rate less than $\approx10^{-11}M_{\odot}$. Exact cutoff is defined in \citet{Behroozi19}.}; (iv) correlation functions for all, quenched, and star-forming galaxies; and (v) measurements of the environmental dependence of central galaxy quenching, using isolation criteria to identify centrals and a counts-in-cylinders-based quantification of $\sim5$~Mpc density. Orphan galaxies are necessary to correct artificially disrupted subhalos in the simulations and are added by extrapolating the positions and velocities of disrupted subhalos \citep{jiang_vdB14}.

\subsection{Dataset}
\label{sec:data}
The positions and velocities of galaxies/halos are defined relative to the position and velocity of the central halo around which these galaxies/halos exist, i.e., $x_i, v_i = x_{i,\rm sub}, v_{i,\rm sub} - x_{i,\rm cen}, v_{i,\rm cen}$. The three-dimensional radius is $r=\sqrt{x^2+y^2+z^2}$.
For projected data, we select the $z$ axis of the simulation box as the line-of-sight. 
The projected radial distance between the cluster's central galaxy and any other galaxy is $R=\sqrt{x^2+y^2}$.
The relative line-of-sight velocity (LOS) of a simulated galaxy is
\begin{equation}
v_{\rm LOS} = (v_{z}-v_{z,{\rm cen}})+aH(z)d_{\rm com,LOS},
\end{equation}
where $v_{z}-v_{z,{\rm cen}}$ is the physical velocity along LOS of the galaxy with respect to the cluster, and $d_{\rm com, LOS}$ is the comoving distance between the cluster and the galaxy along the LOS. We apply a maximum velocity cut $|v_{\rm LOS}| < 3000\ {\rm km\,s}^{-1}$ for the galaxies used to study the velocity distribution. The velocity cut is $\approx [2.5-4]v_c$ for the most massive and least massive clusters.  We normalize all positional information by $r_{\rm 200m}$ and the velocity information by circular velocity, $v_c = \sqrt{GM_{200m}/r_{200m}}$. In total, $1.8\times 10^6$ galaxy-halo pairs are used for the 3D dataset and $5\times 10^6$ galaxy-halo pairs for the 2D dataset.

\begin{figure}[ht]
	\includegraphics[width=0.5\textwidth]{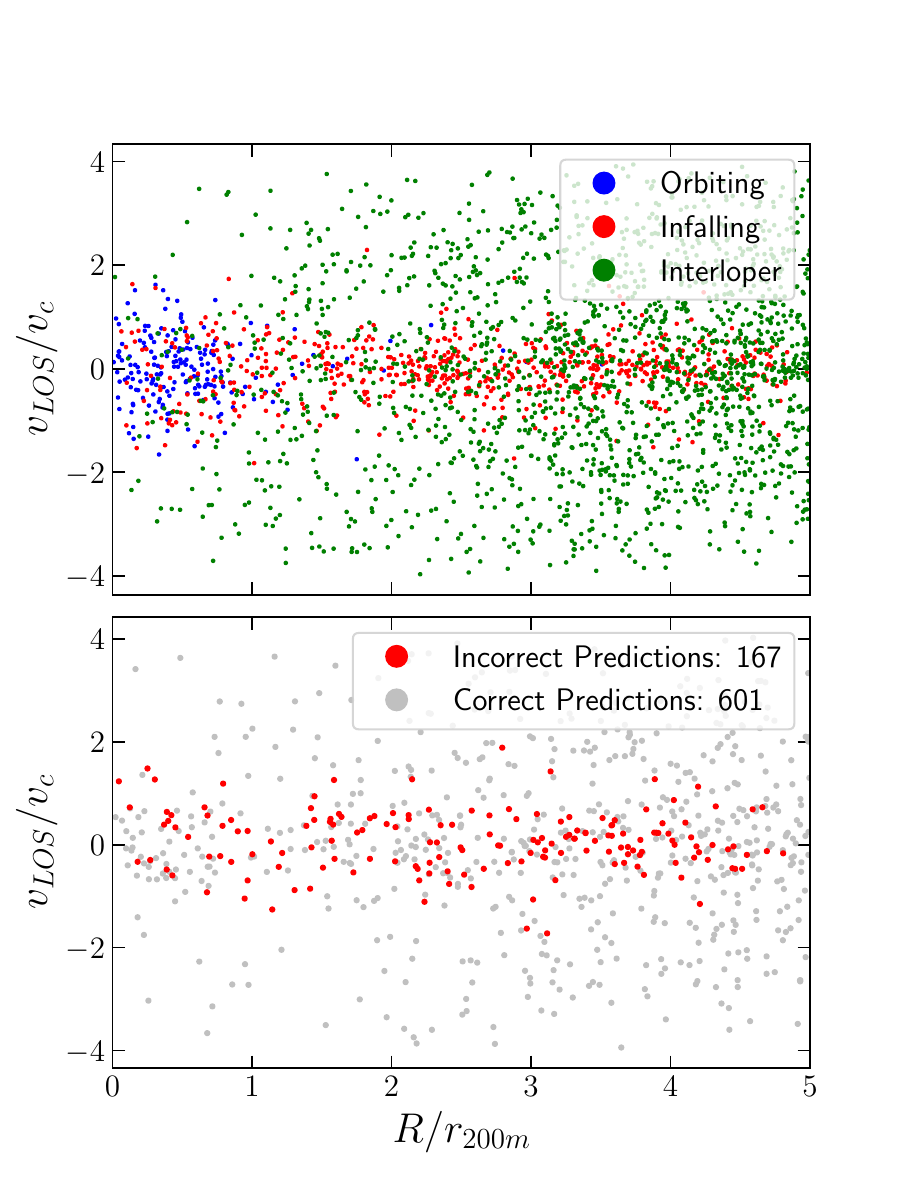}
    \caption{\textit{Top panel}: distribution of the galaxies in line-of-sight velocity $\vlos$ and projected radius $R$, which shows visually how the galaxies are arranged in and around the dark matter halo. The blue dots represent orbiting galaxies, while the red dots represent infalling galaxies. In the 2D distribution, the green dots represent interlopers. \textit{Bottom panel}: distribution of ML predictions using RF showing the distribution of correct and incorrect predictions when applying deterministic classification. The ML correctly predicts the galaxies where there is a clear separation between the populations in the phase space and incorrectly predicts when the populations are mixed. The figure is downsampled from journal version.}
    \label{fig:vr_r}
\end{figure}

\subsection{Classification of Galaxies} 
\label{sec:classification}
We define the three distinct populations of galaxies following \citet{Aung22}: background (when they are outside the turnaround radius), infalling (when they are inside the turnaround radius before they reach their first pericenter), and orbiting (after they have passed the first pericenter). Turnaround radius is measured as the radius at which stacked profile of average radial velocity squared is minimum \citep[equivalent to where average radial velocity is 0, but cleaner, ][]{Pavlidou2014,Korkidis2020}. 
The infalling galaxies generally have larger radii and smaller velocities than the orbiting galaxies, while we expect orbiting galaxies close to the central halo with inward and outward motions. Infalling galaxies are moving toward the inner region of clusters with a large negative radial velocity. Outside the turnaround radius, galaxies are too far from the central halo and generally move away due to the universe's expansion. The top panel of \Cref{fig:vr_r} shows the distribution of these three populations in the projected phase space. We do not exclude merging clusters, and they follow the same categorization as other clusters. For any satellite galaxies that are not part of the central halo, their classification follows that of the immediate host, i.e. if a galaxy is within the subhalo `a' of the halo `b' around a central host `c', `a' is classified as orbiting for `c' if `b' is orbiting, and `a' is infalling if 'b' is infalling.

Compared to \citet{delosRios2021}, which classifies galaxies into 5 classes, we will use 3 galaxy types (orbiting, infalling, and interlopers). In our scheme, orbiting galaxies combine cluster and backsplash galaxies, and infalling galaxies are the combination of infall and recently infalled galaxies in their study.
\citet{delosRios2021} differentiates cluster vs. backsplash and infall vs. recently infalled based on the radial cut at $r_{\rm 200c}$. However, they are not dynamically distinct populations other than having a radial cut, and the radius of $r_{\rm 200c}$ (or $r_{\rm vir}, \r200m$) does not constitute a natural boundary of halo, as galaxies that are orbiting the clusters can still be found outside of $r_{\rm 200c}$ \citep{Aung2021,Diemer2022a}. Thus, we do not adopt the "virial radius cut" in this work. Instead, we study the radial distribution of orbiting and infalling galaxies as a function of the halo-centric radius. 

\section{Property Estimation in the Absence of Perfect Information}
\label{sec:ML} 

Our primary goal is to get an unbiased estimation of the physical properties of dark matter halos, such as the radial number density and velocity dispersion profiles of different types (e.g., infalling and orbiting) galaxies in and around clusters. In simulations, where we have perfect knowledge of galaxy type, estimating cluster properties is straightforward. For example, we will estimate the velocity dispersion of cluster as dispersion of orbiting galaxies. 
However, in practice, galaxy type is unknown and must be estimated. Class estimation problem generally consists of two categories: probabilistic and deterministic. The probabilistic estimation quantifies the likelihood a galaxy belongs to a certain class. On the other hand, the deterministic scheme uses the probability scores, selects a subset of galaxies with a score above the given threshold, and assigns all of them to a certain class. While the latter is often used for science applications, as we will illustrate in Section~\ref{sec:evaluation_distribution}, the deterministic approach gives a biased estimation of the physical quantities, thereby undesired. With the probabilistic approach, however, we can get an unbiased estimation of physical quantities.

\subsection{Classification Schemes}

Probabilistic classifiers output a probability score for each galaxy, whether orbiting, infalling, or background, reflecting the classifier's prediction uncertainty. In this work, we will use two different classification estimation schemes depending on how we use these probability scores to define the class of the individual galaxy.

\subsubsection{Probabilistic Classification}

In this scheme, we use the probabilities for each classification from the ML model as a weight in halo property estimation. This scheme's fundamental assumption is that the probability scores are properly calibrated. For example, if a galaxy is predicted to have a 60\% chance of being infalling, then indeed, 60\% of them should belong to class infalling. With this strong assumption, we use these probabilities as a weight. For instance, the contribution of the above galaxies to the overall distribution of $R/r_{200m}$ for the infalling population is $0.6$. 

Galaxies for which the model is more confident in classifying have more weight in the overall predicted distribution. Traditional metrics,  such as accuracy, recall, and precision, do not apply here, as the galaxy is not predicted as a single class. 

The number of galaxies for the population $c$ is estimated with 
\begin{equation}\label{eq:count_pb}
        n_{c} = \sum_{i=1}^N p_{c,i},
\end{equation}
where $c$ is orbiting, infalling, or background galaxy population, and $p_{c,i}$ is the probability (or weight) of each galaxy $i$ to be classified as $c$. Similarly, the mean line-of-sight velocity is simply the weighed mean of the line-of-sight velocities, where the weighs are the probabilities. The unbiased estimator of variance is given by 
\begin{equation}\label{eq:dispersion_pb}
        \sigma_{c}^2 = \frac{\sum_i p_{c,i}}{(\sum_i p_{c,i})^2 - \sum_i( p_{c,i})^2}\sum_{i=1}^{N} p_{c,i} v_{i, {\rm LOS}}^2,
    \end{equation}
where all sums are from $i=1$ to $N$, and $v_{i, {\rm LOS}}$ is the relative velocity from the mean line-of-sight velocity. This becomes the normal standard deviation formula when equal weights are applied  \citep[see Equations~(10) and (11) in][]{kllr} if probabilities are properly calibrated.\footnote{For example, if the classification is perfect, all orbiting galaxies have $p_{\rm orb,i}=1$, and all infalling and background galaxies have $p_{\rm orb,i}=0$, the equation becomes the normal standard deviation formula.} Therefore, it is important to assess the calibration of the classifier in addition to other test scores. 

\subsubsection{Deterministic Classification}

In the deterministic classification scheme, the class is given by the class-probability $p_{c,i}$ that is highest among all classes, and those with the same class are treated equally in the measuring cluster properties, ignoring the additional uncertainty information that comes with the probabilities. This is equivalent to setting $p_{c,i}= 0$ or 1 in Equations \eqref{eq:count_pb} and \eqref{eq:dispersion_pb}. For instance, The velocity dispersion of each type of galaxy, $c$ (orbiting, infalling, or background) is 
\begin{equation}\label{eq:dispersion_det}
    \sigma_c^2 = \sum_{i \in \{{\rm Type}~c\} } v_{i, {\rm LOS}}^2/(N_c-1),
\end{equation}
where $c \in \{{\rm infalling,~orbiting,~interloper}\}$, and $N_c$ is total number of galaxies of type $c$.
We present the results based on this deterministic classification as a benchmark. 

\subsection{Machine Learning Models}

To estimate the class probabilities, we consider and use three classifiers: Random Forest (RF), k-Nearest Neighbors (kNN), and Logistic Regression. With \textsc{C$^2$-GaMe}, the user can import the trained models, perform a classification task on a new data set, or train a new model with similar parameter choices.

\subsubsection{Random Forest (RF)}

The basis of the RF algorithm is the decision tree. A decision tree gradually progresses through a set of true/false questions and answers in a tree-like structure to reduce the possible range of outcomes until we are confident enough to make a single prediction. 
For example, the decision tree will ask questions such as "what is the radial position" and "what is the radial velocity" etc. After a series of similar questions, the tree will become confident enough to make a single prediction. If the galaxy is far from the cluster, it will likely be an infalling galaxy. On the other hand, if it is close to the cluster with positive radial velocity, it will only have a slight chance of being an infalling galaxy.

In the RF model, the input features form the basis for creating these questions \citep{RF}. In our problem, the original input features to the model are the positions and velocities of the galaxies. The algorithm aggregates predictions from several of these decision trees to construct an overall prediction that is much more accurate than a prediction from any single tree. Since each tree uses a random subset of features, a random subset of the training data (hence "random forest"), and asks slightly different but related questions, diversity in the forest increases, which leads to more robust overall predictions. The algorithm then aggregates the results of decision trees for each element $i$ such that 
\begin{equation}
    p_{c,i} = \frac{\textrm{number of trees predicting $i$ as $c$}}{\textrm{total number of trees}},\label{eq:probability}
\end{equation}
where $p_{c,i}$ is the probability of being classified as $c$ for the galaxy $i$, where $c$ is either orbiting, infalling or background.
We use 100 decision trees in our model, which is the approximate number of decision trees that achieves the optimal performance and efficiency of the algorithm based on our experimentation.

\subsubsection{k-Nearest Neighbor (KNN)}

KNN algorithm classifies data points based on the Minkowski distance in $n^{\rm th}$ dimensional space (where $n$ is the number of input features). For example, in fiducial case, it is a Euclidean distance between $r/r_{\rm 200m}$ and $v/v_c$. The algorithm compiles the $k$ nearest data points ($k$-nearest neighbors) and uses a majority vote to determine the class for the data point in question \citep{kNN}. For example, if $k = 4$, the classifier finds the 4 closest galaxies in the phase space to the target galaxy. It classifies the target galaxy based on the classifications of those 4 neighbors, where the most highly represented class from those 4 neighbors wins. KNN is simpler than random forest conceptually, but it has some limitations. For example, since KNN determines "neighborhoods," there must be some meaningful way to measure the distance between the input data points. This implies that the data features must have a similar order of magnitude, such that the distance metric represents the significant distance between neighbors of different classes.

It can also make a probabilistic classification. The probability of a galaxy $i$ classified as $c$ is
\begin{equation}
    p_{c,i} = \frac{\textrm{number of neighbors around $i$ classified as $c$}}{\textrm{total number of neighbors}}.
\end{equation}
Additionally, the neighbors are weighted according to their distance to the point in question (closer neighbors will have more weight). Based on our experiment, we used a k-value of $k = 15$, which achieves optimal performance without sacrificing computational efficiency.

\subsubsection{Logistic Regression}
Logistic Regression is a statistical method for classification problems that aims to predict the probability of a given target class across multiple classes. Logistic Regression maps input features to a high-dimensional space to model the relationship between the features and the target variable using the logistic function. In binary classification, the probability of the galaxy $x_i$ being positive is expressed by the logistic function:
\begin{equation}
p_i = \frac{1}{1 + {\rm exp}(-\beta^T x_i)},
\end{equation}
where $\beta$ is the coefficient vector,  and $x_i$ is the input vector of the $i^{th}$ galaxy. The coefficient vector $\beta$ is estimated using maximum likelihood estimation, which involves finding the values of $\beta$ that maximize the likelihood of the observed data given the model. 

The method can be extended to the multinomial logistic regression for multiclass classification. In the multinomial case, the predicted class probabilities are given by:
\begin{equation}
p_{c,i} = \frac{\text{exp}(\beta_{c}^T x_i)}{\sum_{k=1}^{K} \text{exp}(\beta_{k}^T x_i)},
\end{equation}
where $K$ is the number of classes, $x_i$ is the input vector of the galaxy $i$, $\beta_{k}$ is the coefficient vector for class $k$, and $p_{c,i}$ is the probability of the galaxy $x_i$ being classified as class $c$.

\section{Evaluation Methods} 
\label{sec:evaluation_methods} 

Finally, to evaluate the performance of our ML models, we consider three sets of evaluation methods: (1) statistical performance, (2) physical verification, and (3) robustness validation. We use statistical methods, including the Receiver operating characteristic (ROC) curve, and  Brier Score (BS), to evaluate the model's performance. However, to ensure our results are physical, we use a set of physical quantities to test and verify the applicability of our ML models for science applications. Finally, we test the robustness of our results to variations in training data by testing the ML model calibrated based on one simulation using another set of simulations. 

\subsection{Statistical Performance} 
\label{sec:statistics}

Metrics constructed based on the confusion matrix are often used to measure and compare the performance of classifiers with deterministic class assignments. These metrics include accuracy, precision, and recall. However, these metrics are unsuitable when we make probabilistic assignments. Instead, we use statistical methods that take probability as an input for probabilistic assignment evaluation. These measures include the area under the ROC curve and Brier Score. As these tests are developed to test binary classifications, we will provide these scores for the classification of orbiting vs. non-orbiting and infalling vs. non-infalling to test the model predictions for orbiting and infalling populations separately.

\paragraph{Accuracy, Precision, and Recall}

We use accuracy, precision, and recall to quantify the performance of deterministic classification. 
Accuracy describes the fraction of correctly classified galaxies. It is the ratio of (TP + TN) / (P + N) where TP (TN) is the number of true positives (true negative), and P (N) is the number of positives (negative) cases, here either infalling or orbiting populations. The precision is the TP / (TP + FP) ratio, where FP is the number of false positive cases. The precision describes the purity of the positive class. Conversely, recall is the TP / (TP + FN) ratio, where FN is the number of false negatives. The recall describes the performance of the classifier in finding positive cases.

\paragraph{Receiver operating characteristic (ROC)}

A ROC curve illustrates the performance of a binary classifier system as its discrimination threshold is varied. It is created by plotting the fraction of true positives out of the positives vs. the fraction of false positives out of the negatives at various threshold settings. Additionally, we report the Area under the ROC Curve (AUC).

\paragraph{Brier Score (BS)}

The BS is a proper scoring rule that measures the accuracy of probabilistic predictions. It applies to tasks in which predictions must assign probabilities to mutually exclusive discrete outcomes. This function returns the mean squared error of the actual outcome $y \in \{0,1\}$ and the predicted probability estimate  $p$ defined by
\begin{equation}
    {\rm BS} = \frac{1}{N_{\text{samples}}} \sum_{i=1}^{N_{\text{samples}}}(y_i - p(x_i))^2 ,
\end{equation}
where the lower BS indicates the more accurate prediction. Although the BS and the AUC provide a valuable diagnostic for the model comparison, informing which model makes a better prediction, these scores cannot determine if a model prediction is trustworthy to obtain an unbiased estimation of the physical quantities. 

\paragraph{Calibration Evaluation}

The probability output of a classifier gives an estimation of confidence in the prediction. Well-calibrated classifiers are probabilistic classifiers for which the probability outputs can be directly interpreted as a confidence level. Equations \eqref{eq:count_pb} and \eqref{eq:dispersion_pb} are unbiased estimators of the number count and variance of class $c$ if the probabilities are well-calibrated. Hence, it is essential to hypothesis test a classifier's calibration and evaluates its performance. We demonstrate the calibration of our classifier via the calibration plots. The calibration plot shows the true frequency of the positive label against its predicted probability for binned predictions.

\subsection{Evaluating Model Classification} \label{sec:evaluation_distribution}

Since our goal is to estimate the physical properties of dark matter halos, evaluating how well the method recovers the true quantities is necessary. This can be done by comparing the distribution of predicted values against the distribution of the corresponding actual values. The better the match between these two distributions, the better the model is at separating different classes of galaxies. We compare the distribution of velocity and position for a given class of galaxies against the distribution of the same predicted class. Namely, we plot the distribution of $r/r_{200m}$ and $v/v_c$ for all galaxies classified as orbiting by the ML model and compare this to the orbiting galaxies classified in simulations, and repeat the process for infalling galaxies. 

Finally, we want to estimate how reliable our model is for potential science applications. As velocity dispersion is a reliable mass proxy, we evaluate our model by measuring how well we can predict the velocity dispersion of the individual clusters. We estimate the true orbiting and infalling dispersion based on the true classification and the predicted velocity dispersion of the galaxy clusters following \Cref{eq:dispersion_pb} using the predicted probabilities from the model.

\subsection{Verification with IllustrisTNG} 

To ensure our model performance when applying to other galaxy catalogs generated from hydrodynamic simulations, we will utilize the galaxy catalog from TNG hydrodynamic simulations. TNG simulation is run with the moving mesh code \texttt{AREPO} \citep{Springel2010EMesh} and includes a full magneto-hydrodynamics treatment with galaxy formation models, as detailed in \citet{Weinberger2017Methods, Pillepich2018Methods}. Haloes are identified via a friends-of-friends (FoF) algorithm, and subhaloes via the \textsc{SubFind} algorithm \citep{Springel2001Subfind, Dolag2009Subfind}, and the merger tree is constructed with \textsc{SubLink} algorithm \citep{sublink}. We use the same galaxy catalog as \citet{Anbajagane2022} from TNG300, which has a volume of $302.6 {\rm Mpc}^3$, and generate a projected mock catalog following \Cref{sec:data,sec:classification}.

\section{Results}
\label{sec:results}

We investigate how well the deterministic and probabilistic classification schemes work using the realistic 2D projected data from mock galaxy catalogs, where the normalized line-of-sight velocity $\vlos/v_c$ and projected radius $R/\r200m$ are used as input features. We show that the radial distributions of different types of galaxies are well reproduced when using probabilistic classification. We evaluate how well we can predict the velocity dispersion of galaxy clusters based on the classification algorithm. We then show how additional features, such as specific star formation rates, improve the classification. Finally, we test our ML classification technique (trained using the MDPL2+UM mock galaxy catalog) using Illustris-TNG hydrodynamical simulations. 

\subsection{Statistical Performance}
\begin{figure}[ht]
    \centering
    \includegraphics[width=0.49\textwidth]{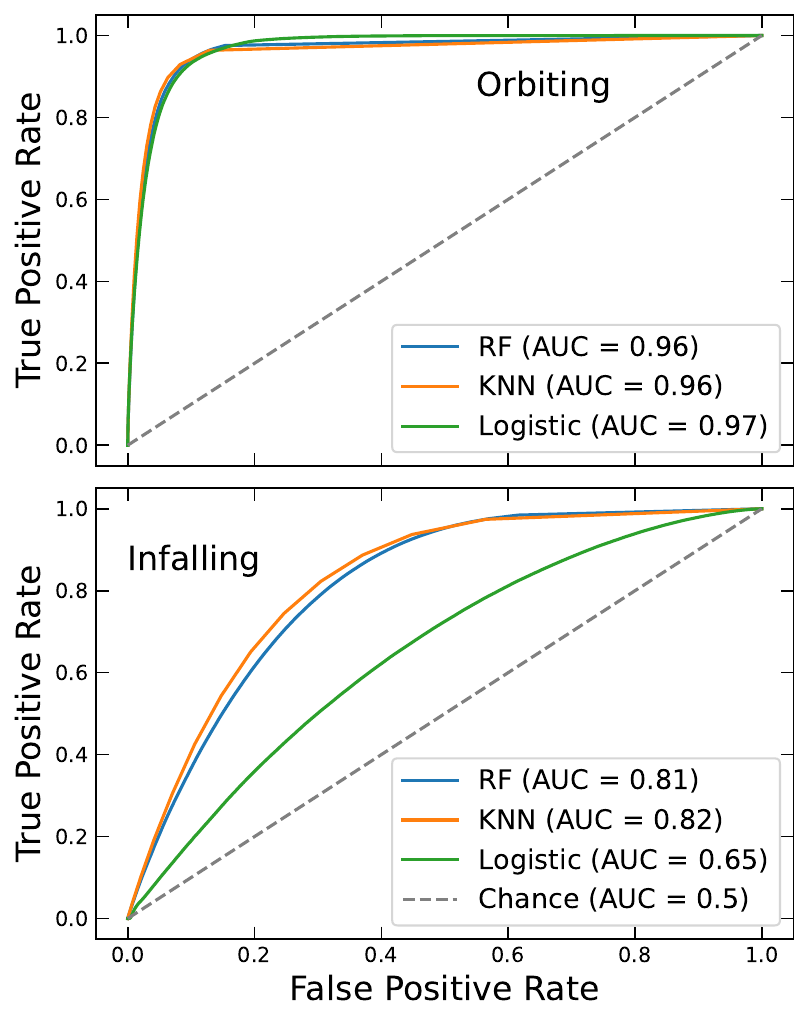}
    \caption{ROC curve for classification of orbiting vs non-orbiting galaxies \textit{(top panel)} and infalling vs non-infalling galaxies \textit{(bottom panel)}. The closer the ROC is to a step function at 0, the closer AUC is to 1, and the more accurate the classification is.}
    \label{fig:roc}
\end{figure}

\begin{figure}[ht]
    \centering
    \includegraphics[width=0.49\textwidth]{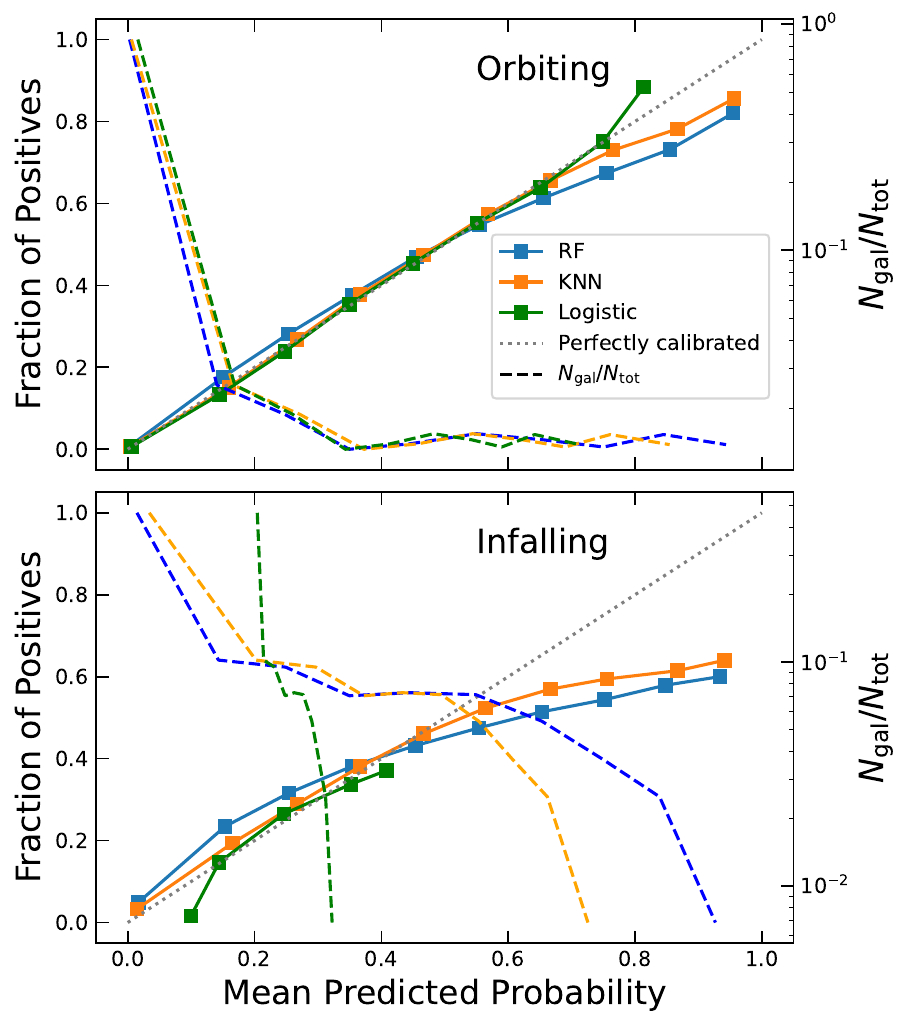}
    \caption{Calibration plot for orbiting vs non-orbiting galaxies \textit{(top panel)} and infalling vs non-infalling galaxies \textit{(bottom panel)}. The x-axis of calibration shows the mean predicted probability binned together, and the y-axis shows the fraction of positives in those bins. The closer the calibration curve is to the line of $y=x$, the better the classifier is calibrated. We also show the number of galaxies in each bin $N_{\rm gal}$ divided by the total number of the type of galaxies $N_{\rm tot}$. The smaller the number of galaxies, the less important the bin is for perfect calibration.}
    \label{fig:calibration}
\end{figure}
We classify the galaxies from the projected mock galaxy catalog with normalized line-of-sight velocity and projected radius as input features to classify the galaxies into one of the orbiting, infalling, and background (interloper) populations. The performance scores are listed separately in Table~\ref{tab:scores} for different classifiers for binary classification of orbiting vs. non-orbiting and infalling vs. non-infalling populations. All classifiers perform better for classifying orbiting vs. infalling populations with better accuracy, precision, and recall. This is because the orbiting population is confined to a low velocity and radius region where it can only be mistaken for an infalling population. In contrast, an infalling population can be mistaken as either orbiting or interlopers, depending on their position in the phase space. This is highlighted in the bottom panel of \Cref{fig:vr_r}, which shows where the correct and incorrect predictions lie with deterministic classification. The incorrect populations appear where there is no clear separation between different populations, which occurs when infalling galaxies are mixed with either orbiting or interloper populations.

RF and KNN perform similarly between different classifiers, with Logistic Regression performing the worst. \Cref{fig:roc} shows the ROC curve and its corresponding area under the curve (AUC) for different classifiers, showing similar results between all three classifiers. However, statistical metrics involving probabilistic classification -- i.e., ROC, AUC, Brier Score -- RF and KNN, generally perform better than Logistic Regression. \Cref{fig:calibration} shows the calibration plot, showing the mean predicted probability vs. the fraction of orbiting (top panel) or infalling galaxies (bottom panel). The closer the calibration curve is to the line of $y=x$, the better calibrated the classifier is. This again supports that all classifiers are close to perfect calibration. However, they deviate from perfect calibration for infalling galaxies for large predicted probability. This implies that identifying infalling galaxies with high confidence from the classifier is unreliable. However, the number of infalling galaxies in these bins is small and, thus, should not impact our physical predictions, as shown in the following sections.

\begin{table*}[ht]
    \centering
    \begin{tabular}{|c|c|c|c|c|c|c|c|c|c|c|c| }
        \hline
        \multicolumn{2}{|c|}{} &  \multicolumn{5}{|c|}{Orbiting} &  \multicolumn{5}{|c|}{Infalling}\\
        \hline
        \multicolumn{2}{|c|}{} & Acc. & Prec. & Recall & AUC & BS & Acc. & Prec. & Recall & AUC & BS \\
        \hline
        \hline
        \multirow{3}{*}{Fiducial} & RF & 0.947 & 0.646 & 0.698 & 0.961 & 0.036 & 0.773 & 0.512 & 0.460 & 0.811 & 0.1474\\
        & KNN & 0.950 & 0.653 & 0.753 & 0.959 & 0.034 & 0.785 & 0.543 & 0.472 & 0.823 & 0.138 \\
        & Logistic & 0.943 & 0.600 & 0.766 & 0.968 & 0.036 & 0.759 & 0.350 & 0.041 & 0.653 & 0.170 \\
        \hline
        \hline
        \multirow{3}{*}{+mratio} & RF & 0.954 & 0.687 & 0.743 & 0.970 & 0.032 & 0.785 & 0.545 & 0.451 & 0.830 & 0.136 \\
        & KNN & 0.952 & 0.663 & 0.778 & 0.961 & 0.032 & 0.788 & 0.550 & 0.474 & 0.825 & 0.137 \\
        & Logistics & 0.944 & 0.606 & 0.772 & 0.968 & 0.036 & 0.761 & 0.372 & 0.044 & 0.654 & 0.1696\\
        \hline
        \hline
        \multirow{3}{*}{+ssfr} & RF & 0.958 & 0.712 & 0.797 & 0.975 & 0.029 & 0.790 & 0.555 & 0.498 & 0.838 & 0.133 \\
        & KNN & 0.954 & 0.676 & 0.774 & 0.962 & 0.031 & 0.787 & 0.549 & 0.479 & 0.827 & 0.137 \\
        & Logistics & 0.949 & 0.636 & 0.780 & 0.972 & 0.034 & 0.760 & 0.384 & 0.054 & 0.660 & 0.169 \\
        \hline
    \end{tabular}
    \caption{Comparison of the performance of different classifiers using various metrics listed in \cref{sec:statistics} using only normalized projected radius and line-of-sight velocity as input features for fiducial case, fiducial with mass ratio, fiducial with mass ratio and specific star formation rate. We abbreviate accuracy as ``Acc.'' and precision as ``Prec.''.}
    \label{tab:scores}
\end{table*}

\subsection{Evaluation with Physical Quantities}
\label{sec:2d}
To verify that we can measure physical quantities from predicted classes, we calculate how well the \textsc{C$^2$-GaMe} models predict the distribution of each type of galaxy in projected distance and velocity. We also compare the prediction for the line-of-sight velocity dispersion, a summary statistic of the distribution useful for inferring cluster mass.

\subsubsection{Projected density and velocity distribution}
\begin{figure*}
	\includegraphics[width=1\textwidth]{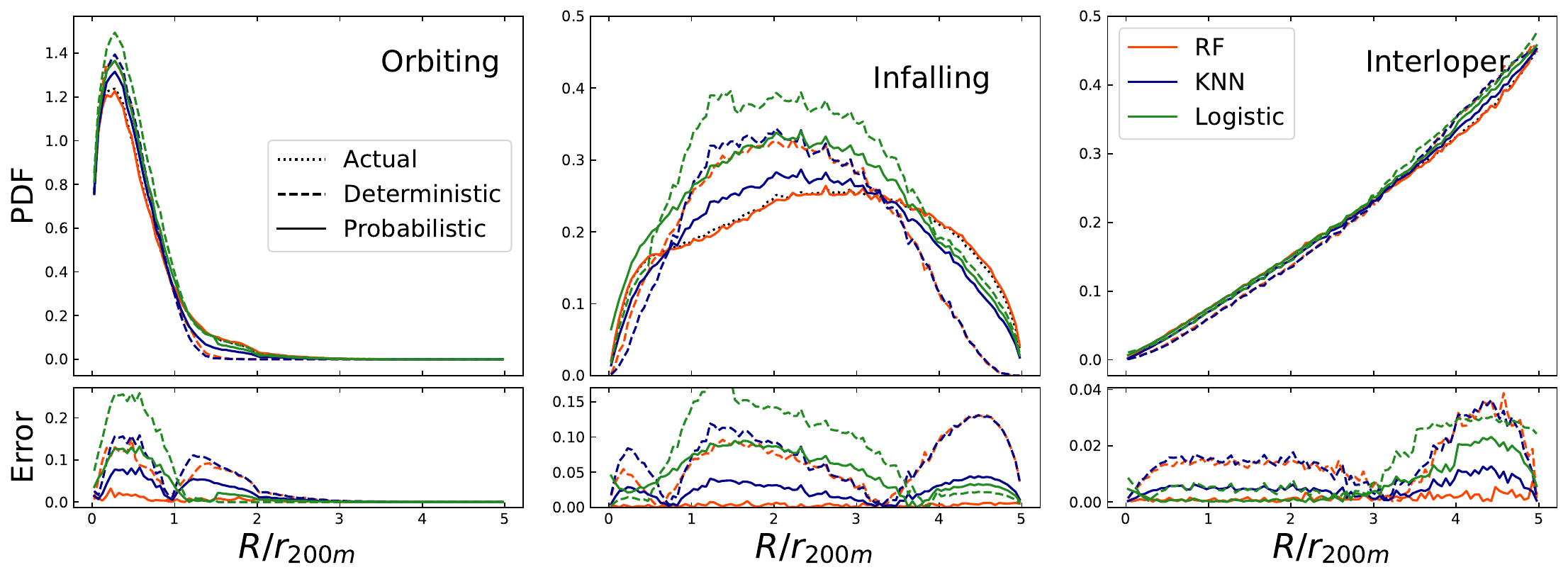}
    \caption{
    Comparison of probability density function as a function of $R$ between deterministic classification prediction (dashed), probabilistic prediction (solid), and actual (black dotted) using Random Forest (orange), KNN (blue), and Logistic (green) classifiers for different galaxy populations using 2D data, showing the probability density function (\textit{top panel}) and the absolute difference between actual and predicted PDF (\textit{bottom panel}). The logistic regression model (green) performs the worst for both probabilistic and deterministic distributions. The probabilistic distribution from the random forest model (orange) reproduces the actual distribution with the smallest error of $<3\%$. We note that the deterministic classification method is similar to the method used in \citet{delosRios2021}.}  
    \label{fig:r0_dists_2d}
\end{figure*}

\begin{figure*}
	\includegraphics[width=0.9\textwidth]{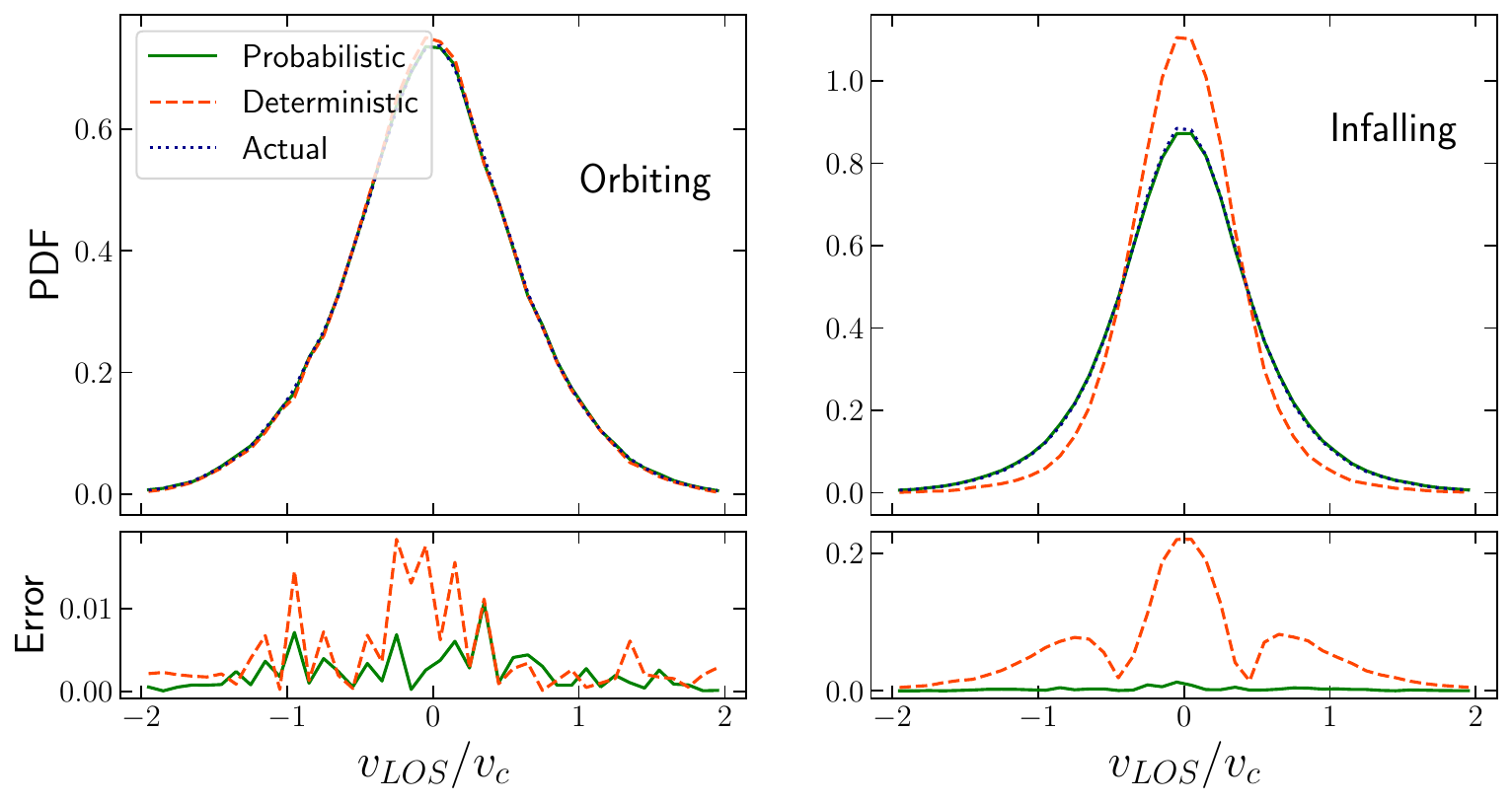}
    \caption{
    The probability density function of line-of-sight velocity $\vlos$ of actual, deterministic-classification, and probabilistic classifications using the orbiting (\textit{left panel}) and infalling (\textit{right panel}) populations, showing the probability density function (\textit{top panel}) and the difference between actual and predicted PDF (\textit{bottom panel}). Unlike radial distributions, we see little difference between deterministic and probabilistic classification when reproducing orbiting velocity distribution, but probabilistic classification still performs slightly better when reproducing infalling velocity distribution.}  
    \label{fig:v_dist_match_infall}
\end{figure*}

To compare the distribution of the galaxies, we will compute the probability density function as a function of projected distance and line-of-sight velocity. For each type of galaxy, we compare how many galaxies are at each normalized projected distance bin (Fig \ref{fig:r0_dists_2d}) or line-of-sight velocity bin (Fig \ref{fig:v_dist_match_infall}) for true dataset vs. prediction based on deterministic or probabilistic classification \Cref{eq:count_pb}.

\Cref{fig:r0_dists_2d} shows the PDF of orbiting, infalling, and background galaxies for probabilistic and deterministic classification schemes using RF, KNN, and Logistic classifiers. 
The deterministic classification shows a relatively poor match between the predicted and actual distributions of $R/r_{200m}$. As mentioned in \Cref{sec:evaluation_methods}, we value the match in distribution more than the scores for individual deterministic classification. The distribution matches become significantly better when using the probabilistic output scheme. This agreement is more potent when using RF compared to KNN and Logistic Regression, where the difference in PDF is less than 0.03 for the RF classifier for all types of galaxies. Compared to KNN, the differences can be more significant than 0.05. 

\begin{figure*}
	\includegraphics[width=0.99\textwidth]{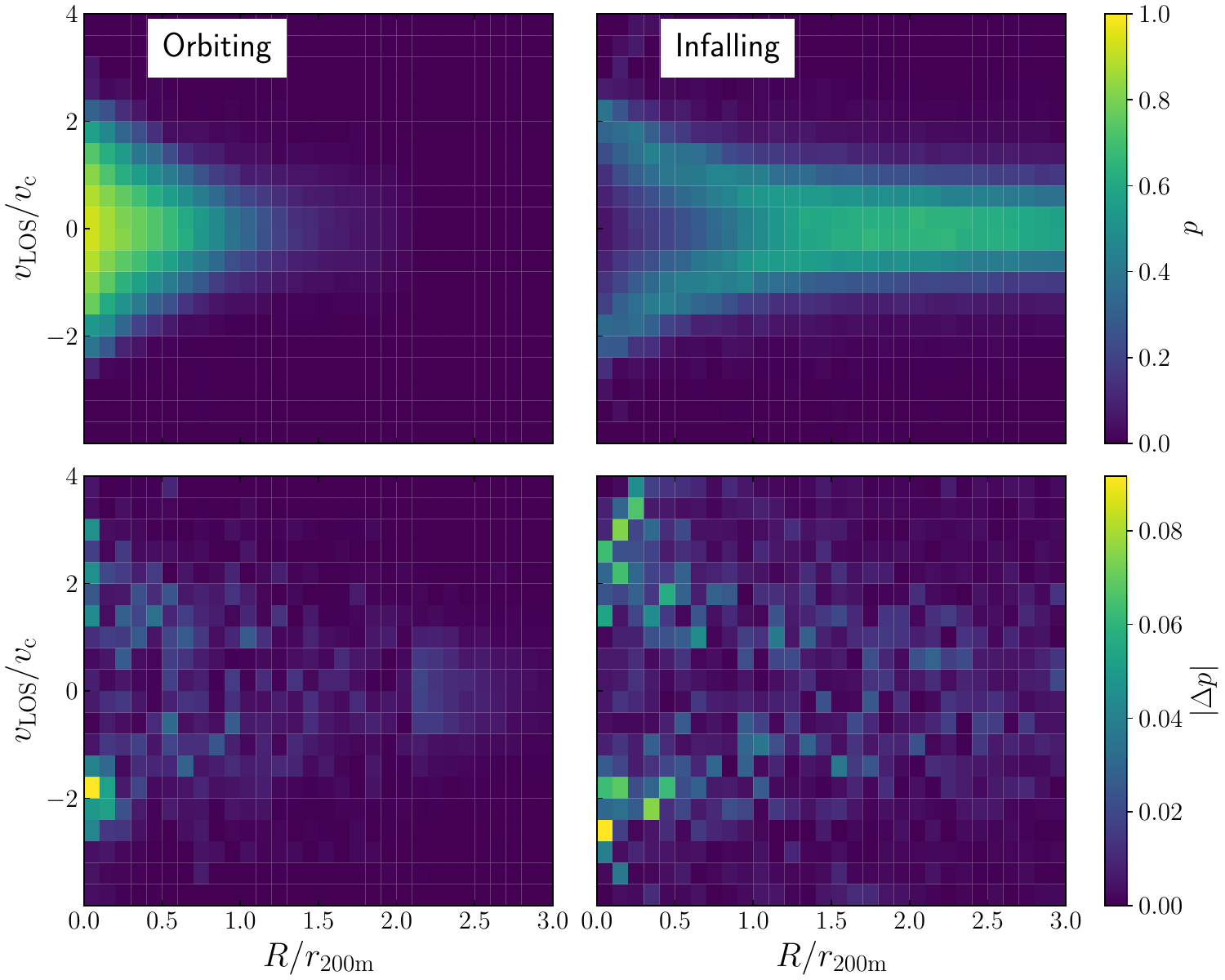}
    \caption{
    {\it Top panels}: Probability of orbiting ({\it left panel}) and infalling ({\it right panel}) as a function of projected radius $R$ and line-of-sight velocity $v_{\rm los}$. Both show symmetric distribution around $v_{\rm los}=0$, where orbiting galaxies are more likely to occur at small $R$ and $v_{\rm los}$. In contrast, the infalling galaxies are more likely to occur at intermediate $R$ and $v_{\rm los}$. The regions at large $R$ and $v_{\rm los}$ are dominated by interlopers (not shown). \textit{Bottom panels} show the difference between the actual and predicted probabilities, where the maximum difference does not exceed $0.1$ when the bin size is 0.1 in $R/r_{\rm 200m}$ and 0.4 in $v/v_c$. This maximum difference depends on the bin size, and increases as bin size decreases and the data gets noisy.}  
    \label{fig:p_dists_2d}
\end{figure*}

Despite similar performance between RF, KNN, and Logistic Regression in statistical metrics and scores, we will use the RF classifier from here on due to the significantly better performance in matching physical distributions. \Cref{fig:p_dists_2d} shows the probability of orbiting or infalling as a function of projected radius and line-of-sight velocity. Even though the two populations are mixed, we see a faint separation between the two populations, where galaxies with small $R$ and $v_{\rm los}$ are more likely to be orbiting. In contrast, galaxies with intermediate $R$ and $v_{\rm los}$ are more likely to be infalling. However, the two populations have no clear separation based on the projected phase space using $R$ and $\vlos$.  

\subsubsection{Line-of-sight  Velocity Dispersion}

\begin{figure}[ht]
	\includegraphics[width=0.5\textwidth]{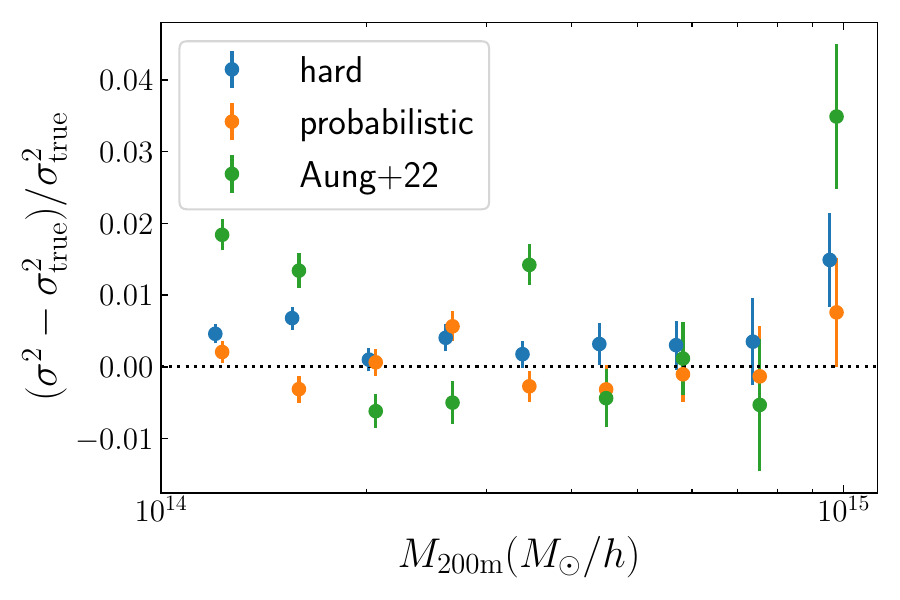}
    \caption{
    The fractional error of predicted orbiting velocity dispersion from RF as a function of cluster mass, where the error bars are the standard error on the mean. Our ML-based classification method produces a dispersion estimate with $\lesssim1\%$ accuracy, while the semi-analytic method based on \citet{Aung22} predicts the velocity dispersion with the accuracy of $\lesssim4\%$.}  
    \label{fig:vd_ratio_mass_dependence}
\end{figure}

We estimate the orbiting and infalling velocity dispersion of the galaxy clusters following \Cref{eq:dispersion_pb}. \Cref{fig:vd_ratio_mass_dependence} shows the fractional error of the predicted velocity dispersion squared of orbiting galaxies for various mass bins and explores the differences in the two RF output methods in calculating velocity dispersion as well as the classification following the physical model based on \citet{Aung22}, where the velocity dispersion of orbiting galaxies is given by $\int v^2\rho_{\rm orb}(r)G(v/\sigma_{\rm orb}) {\rm d}r {\rm d}v / \int \rho_{\rm orb}(r){\rm d}r$, where $G(v/\sigma_{\rm orb})$ is a Gaussian function in the likelihood function in Equation~2 in \citet{Aung22}. When measuring velocity dispersion, we specifically select orbiting galaxies with $R>0.3\r200m$ to avoid the region where the fraction of orphan galaxies is higher. The inner region is also prone to error when using different catalogs (see \Cref{sec:validation}) or when excluding orphans, and the model in \citet{Aung22} is not properly fitted in this region.

The deterministic and probabilistic classification methods provide equally accurate velocity dispersion estimates. Comparatively, the velocity dispersion based on the non-ML technique produces larger fractional errors. The ML-based method predicts the fractional error in the velocity dispersion squared with better than $\lesssim1\%$ accuracy, while the \citet{Aung22} model predicts the velocity dispersion with the accuracy of $\lesssim4\%$. 
The \citet{Aung22} model describes the probabilistic distribution of orbiting and infalling galaxies, which correctly models the phase space where different populations are mixed. If a phase space cut is applied for galaxy classification without taking into account the mixing region, the velocity dispersion of the orbiting (infalling) population is underestimated (overestimated), as all galaxies with small (large) velocity are classified as orbiting (infalling) (see \Cref{fig:p_dists_2d} and \Cref{sec:2d}). 

\subsection{Additional Features}
\label{sec:additional_features}

\begin{figure}[ht]
	\includegraphics[width=0.5\textwidth]{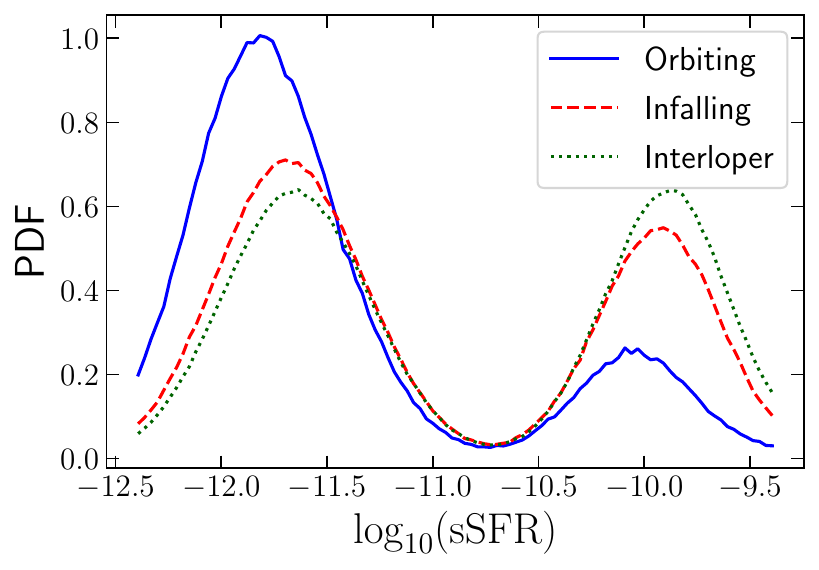}
    \caption{
    This bimodal distribution shows the distinct properties of orbiting vs. infalling galaxies regarding their specific star formation rate (sSFR). As shown in the distribution as a higher peak for orbiting at lower values, orbiting galaxies are more likely to have lower sSFR, because the star formation in these galaxies becomes quenched as they fall towards the halo center. Conversely, infalling galaxies are more likely to have higher star formation rates, evident by the higher peak on the right. Since sSFR can effectively separate these populations, we add it as a feature in the RF model.}  
    \label{fig:bimodal_sfr}
\end{figure}

Other properties of galaxies can also improve the classification of the galaxies. Here, we will explore two of them: the mass ratio of the galaxy stellar mass to host cluster dark matter mass and the specific star formation rate (${\rm sSFR} \equiv {\rm star\,formation\,rate}/M_{*}$) of the galaxies. Galaxies in cluster environments are known to be redder, i.e., they have stopped star forming due to the gas stripping and strangulation in the hot intracluster medium \citep{GunnGott,Abadi99,Larson80}. \Cref{fig:bimodal_sfr} shows that orbiting galaxies are more likely to have smaller star formation rates than the infalling and background galaxies, which are also quenched due to feedback effects \citep{Silk98}. In addition, galaxies with pericentric passage inside the cluster experience dynamical friction and tidal stripping and have their mass slowly decayed over time \citep{jiang_vdB14}. Thus, we add a logarithmic value of sSFR and mass ratio as input features in addition to $R$ and $v_{\rm los}$. 

\Cref{tab:scores} shows the performance statistics after adding mass ratio, and \Cref{tab:scores} shows the performance statistics after adding both mass ratio and sSFR. Adding new features generally improves the classifier results, but most notably, RF has the most improvement and now outperforms all other classifiers. The classification of orbiting population is still significantly better than the infalling population with better improvements. This is because the clusters affect both mass ratio and sSFR, specifically after falling into the clusters. The infalling and interloper populations have similar properties for sSFR and mass ratio (see \Cref{fig:bimodal_sfr}); thus, no additional information is gained in differentiating these two populations.

\subsection{Robustness Verification with Different Galaxy Catalogs}
\label{sec:validation}
While the classification algorithm works well when tested on the set generated from the same simulation as the training set, we provide the validity of applying our models to other galaxy catalogs generated from hydrodynamic simulations. We will validate our model with catalogs from TNG hydrodynamic simulations, which are not used in training or test datasets. \Cref{tab:scores_tng}  summarizes performance scores for various statistics when applying our model trained on MDPL2 data, and using only radius and velocity as input features, to the TNG dataset. The model's performance is significantly lower than when applying this model to the MDPL2 dataset. The performance degradation is similar when including mass ratio as a feature (\Cref{tab:scores_tng}). However, the performance further decreases when also including sSFR as an input feature (\Cref{tab:scores_tng}), where the accuracy for the orbiting population decreases to around $81\%$ accuracy (a decrease of around 7\%), and to $53\%$ accuracy for the infalling population (a decrease of around only 3\%). This implies that the difference in TNG and our original catalog of MDPL2+UM differs mainly in orbiting population. For further insights into the performance and physical interpretations, we will interpret physical metrics such as the PDF of the projected radius.  

\begin{figure}[ht]
	\includegraphics[width=0.49\textwidth]{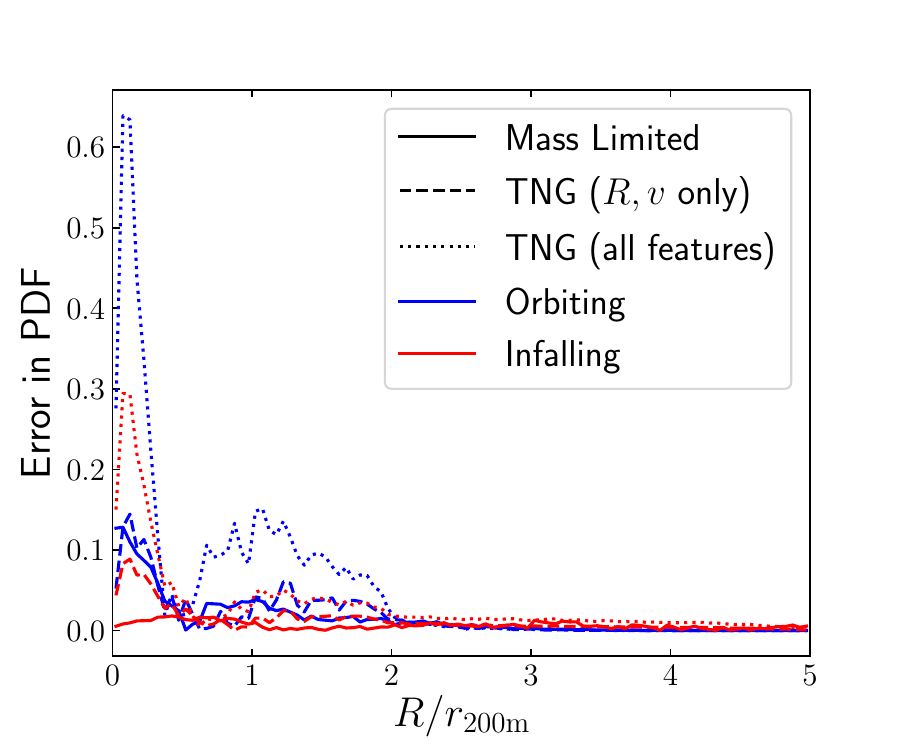}
    \caption{
    Absolute error in the probability distribution function of orbiting ({\it blue}) and infalling ({\it red}) galaxies. The error is calculated between the actual distribution and that predicted from probabilistic classification. For the mass-limited sample, the predicted distributions are generated using an RF model trained on MDPL2 data with a mass cutoff of host mass $>10^{14.2}\Msunh$, and tested on the original sample, which has a mass cutoff of host mass $>10^{14}\Msunh$. For TNG, we either use all features tested in MDPL2 or limit the features to only projected radius $R$ and line-of-sight velocity $\vlos$. }  
    \label{fig:validation}
\end{figure}

\Cref{fig:validation} shows the distribution match for orbiting and infalling galaxies for different validation tests using probabilistic classification. For the model that only employs position and velocity information, the error in PDF stays within $15\%$ inside $0.5\r200m$ and is less than $5\%$ outside. This is either because of the baryonic physics, which introduces differences in the inner region of the cluster \citep[e.g., velocity bias, ][]{wu13}, or different disruption rates and orphan treatment in the different simulations. In fact, for comparison, we also train the algorithm using the original cluster sample (MDPL2+UM) with $M_{\rm 200m}>10^{14.2}\Msunh$, and test it on the original sample with $M_{\rm 200m}<10^{14.2}\Msunh$. The error for using different simulations is roughly comparable to applying the model to a mass-limited test sample where we test with galaxies around clusters with masses smaller than the trained sample. 

When we also compare the predicted and actual probability distribution functions when applying the model that includes a specific star formation rate as an additional feature, we found that the discrepancy can reach up to $60\%$ in the inner regions. This is expected as the star formation rate distribution in the orbiting and the infalling galaxies in TNG are slightly different than that of the MDPL2+UM catalog. Thus, an algorithm trained with the star formation rate on UM catalog will not produce an accurate result when using the star formation rate from TNG. MDPL2+UM catalog is fitted to reproduce star formation rate across different redshifts. Hence, the algorithm trained on the catalog should apply to observational data but may not apply to simulations with slightly different star formation rates. However, further work on the error associated with using sSFR is warranted by contrasting with other simulations of similar distribution. In addition, obtaining the mass ratio in the observations is harder, where dark matter halo mass needs to be estimated through weak lensing or other proxies, making the model with mass ratio harder to apply in observations. We note the excellent agreement on different simulation datasets with just projected radial distance and line-of-sight velocity. 

\begin{table*}[ht]
    \centering
    \begin{tabular}{|c|c|c|c|c|c|c|c|c|c|c|c| }
        \hline
        \multicolumn{2}{|c|}{} &  \multicolumn{5}{|c|}{Orbiting} &  \multicolumn{5}{|c|}{Infalling}\\
        \hline
        \multicolumn{2}{|c|}{} & Acc. & Prec. & Recall & AUC & BS & Acc. & Prec. & Recall & AUC & BS \\
        \hline
        \hline
        \multirow{3}{*}{Fiducial} & RF & 0.955 & 0.718 & 0.820 & 0.981 & 0.031 & 0.805 & 0.601 & 0.455 & 0.849 & 0.131 \\
        & KNN & 0.952 & 0.689 & 0.840 & 0.967 & 0.033 & 0.804 & 0.599 & 0.447 & 0.812 & 0.137 \\
        & Logistics & 0.949 & 0.662 & 0.887 & 0.982 & 0.032 & 0.763 & 0.365 & 0.040 & 0.671 & 0.166 \\
        \hline
        \hline
        \multirow{3}{*}{+mratio} & RF & 0.956 & 0.703 & 0.886 & 0.985 & 0.028 & 0.815 & 0.641 & 0.447 & 0.864 & 0.123 \\
        & KNN & 0.953 & 0.685 & 0.885 & 0.973 & 0.031 & 0.807 & 0.609 & 0.444 & 0.813 & 0.137 \\
        & Logistics & 0.950 & 0.660 & 0.899 & 0.983 & 0.031 & 0.764 & 0.371 & 0.037 & 0.672 & 0.167 \\
        \hline
        \hline
        \multirow{3}{*}{+ssfr} & RF & 0.949 & 0.704 & 0.738 & 0.981 & 0.033 & 0.803 & 0.585 & 0.493 & 0.850 & 0.129 \\
        & KNN & 0.957 & 0.718 & 0.863 & 0.975 & 0.028 & 0.807 & 0.604 & 0.464 & 0.818 & 0.135 \\
        & Logistics & 0.950 & 0.655 & 0.941 & 0.987 & 0.028 & 0.763 & 0.398 & 0.063 & 0.675 & 0.166 \\
        \hline
    \end{tabular}
    \caption{Comparison of the performance of different classifiers using various metrics listed in \cref{sec:statistics}. The models were trained using galaxy catalogs from the MDPL2 simulation and tested on galaxy catalogs from the TNG simulation. We use only normalized projected radius and line-of-sight velocity as input features for fiducial case, fiducial with mass ratio, fiducial with mass ratio and specific star formation rate. We abbreviate accuracy as ``Acc.'' and precision as ``Prec.''.}
    \label{tab:scores_tng}
\end{table*}

\section{Science Applications} \label{sec:application} 

The classification of orbiting and infalling galaxies around galaxy clusters is important in deriving meaningful cosmological and astrophysical measurements from galaxy clusters. For example, line-of-sight projection mixes different classes in phase space, which need to be separated for dynamical mass estimation. Galaxies that have gone through pericenters and orbiting the clusters experience quenching and can splashback to outer radii. Here, we will focus on dynamical mass estimation as an example application for our method.

\begin{figure}[ht]
    \centering
    \includegraphics[width=0.49\textwidth]{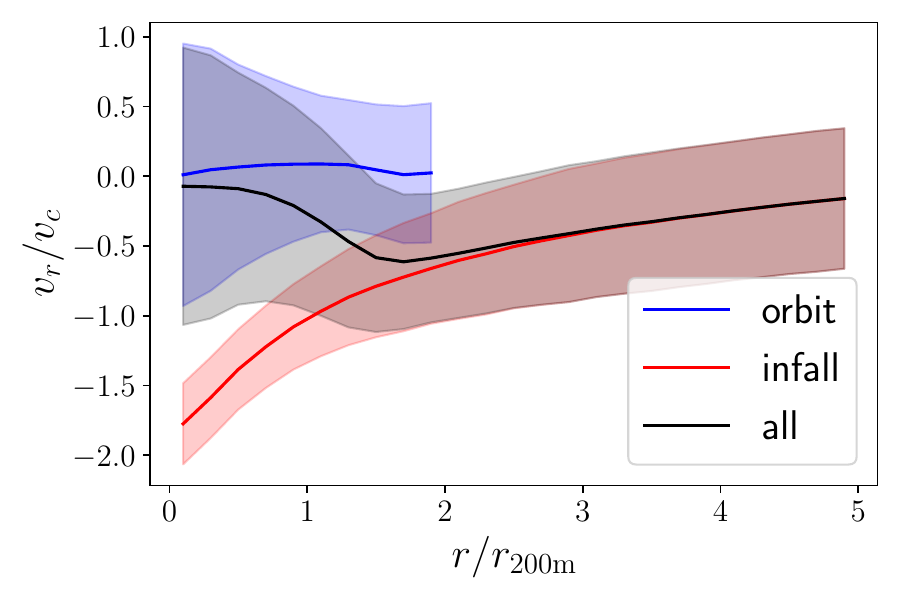}
    \caption{Peculiar radial velocity profile of all ({\it black}), orbiting ({\it blue}) and infalling ({\it red}) galaxies around dark matter halos, where the line indicates the mean and the shaded band indicates the standard deviation. The orbiting galaxies have $\ave{v_r}\approx0$, indicating they are virialized. The infall stream has negative radial velocity, and its magnitude increases toward the halo center.}
    \label{fig:radialv}
\end{figure}

The traditional dynamical mass estimation method relies on linking the density and mass profile of the dark matter halo to the velocity dispersion through the Jeans equation derived from the Boltzmann equation. The method usually makes simplified assumptions for average radial velocity and velocity anisotropy  \citep{Wojtak07,Mamon2013}, which can bias the mass estimation if not modeled correctly. However, the relation is only valid when only one population has the same average velocity and velocity dispersion. If we select all galaxies within a certain radius (e.g., $r_{\rm 200m}$), we can identify two distinct populations with different average radial velocities \citep{Aung2021}. \Cref{fig:radialv} shows the average and standard deviation of the radial velocity of different types of galaxies. Thus, the galaxies around the galaxy clusters can be better thought of as two different populations in the Boltzmann equation, where each population has its average velocity and velocity dispersion to balance the gravitational potential of the dark matter halo.

\begin{figure}[h]
    \centering
    \includegraphics[width=0.49\textwidth]{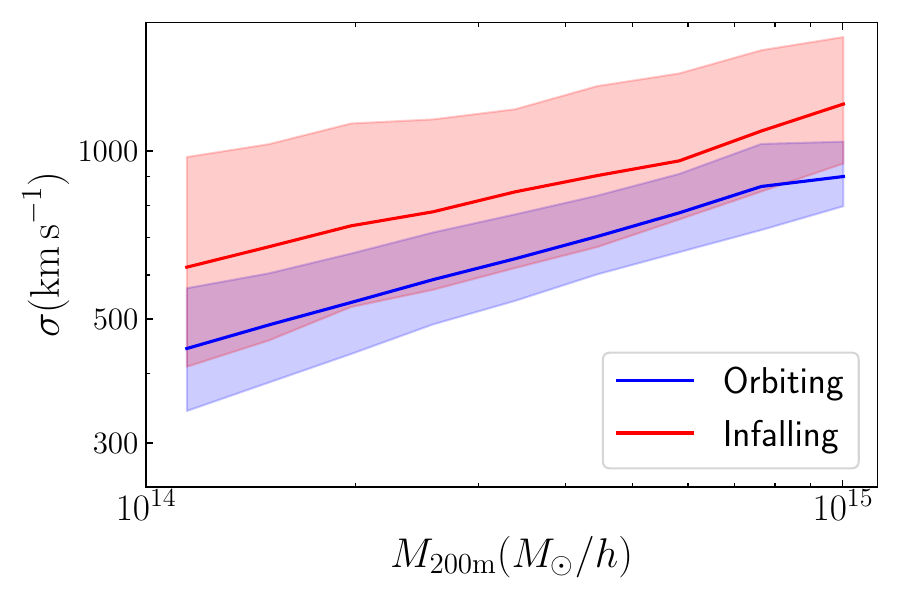}
    \caption{The velocity dispersion-mass relation of orbiting ({\it blue}) and infalling ({\it red}) galaxies. The line indicates the median, and the shaded region shows the 16-84 percentiles of the velocity dispersion of galaxy clusters as a function of cluster mass. Both orbiting and infalling galaxies provide independent tracers of cluster mass.}
    \label{fig:disp_mass}
\end{figure}
In addition, we can obtain independent dynamical mass estimates from different tracers by classifying the galaxies into different types. This is indicated in \Cref{fig:disp_mass} as velocity dispersions of both populations scale proportionally with mass. Limiting the population to orbiting galaxies only in the inner region of the cluster removes the velocity anisotropy introduced by strong radial inflow from infalling galaxies. On the other hand, limiting to infalling galaxies only reduces systematics such as dynamical friction and baryonic effects, which introduce velocity bias to the orbiting galaxies and improve the statistics by allowing us to use the information of galaxies far away from the cluster center. 
\begin{table}[ht]
    \centering
    \begin{tabular}{|c|c|c|c|}
        \hline
        Galaxies & Classification & $a$ & $b$  \\
        \hline
        \multirow{3}{*}{Orbiting} & Actual & 0.38 $\pm$ 0.03 & 2.61 $\pm$ 0.04 \\
        & Deterministic & 0.38 $\pm$ 0.04 & 2.60 $\pm$ 0.07 \\
        & Probabilistic & 0.37 $\pm$ 0.03 & 2.61 $\pm$ 0.04 \\
        \hline
        \multirow{3}{*}{Infalling} & Actual & 0.28 $\pm$ 0.03 & 2.55 $\pm$ 0.04 \\
        & Deterministic & 0.28 $\pm$ 0.03 & 2.54 $\pm$ 0.06 \\
        & Probabilistic & 0.29 $\pm$ 0.03 & 2.57 $\pm$ 0.05 \\
        \hline
    \end{tabular}
    \caption{Best-fits parameters of the linear model for $\log_{10} \sigma$ and $\log_{10}M$ in \Cref{fig:disp_mass} and \Cref{eq:dispersion_pb}. }
    \label{tab:my_label}
\end{table}

To estimate the predictability of the velocity dispersion, we fit both the actual and predicted dispersion vs. mass with a log-linear relation,
\begin{equation}
    \log_{10} \sigma = a \log_{10} \left( \frac{M}{10^{14} M_{\odot}/h} \right)+ b, 
\end{equation}
and the parameters of the fit are listed in \Cref{tab:my_label}. The scatter $\ave{\delta^2}^{1/2}$ is $\approx 0.1$ for orbiting population and $\approx0.15$ for infalling populations.
The dispersion predicted follows the same slope, intercept, and scatter within $1.5\%$ of the actual galaxies. Our results show that the velocity dispersion-mass relation calibrated from the orbiting galaxies, as classified by the RF algorithm, follows the same relation as the actual orbiting galaxies in the simulation. The algorithm achieves this resemblance to the actual data by successfully removing the contributions from interlopers and infalling galaxies.

We note that our algorithm works on normalized distance and velocities instead of physical units, while we want to apply the classification algorithm before the mass estimation. Thus, one may need some estimate of the radius and mass of the cluster, through some simple scaling relations such as cluster richness or X-ray brightness. Then we can apply the classification algorithm, get a dynamical mass estimate and iterate the process in order to obtain a more accurate prediction.  

\section{Conclusions}
\label{sec:conc}

This paper presents \textsc{C$^2$-GaMe}, a suite of machine learning (ML) models for classifying cluster galaxies into orbiting, infalling, and background interloper populations using position and velocity information in projected phase space. The primary goal of \textsc{C$^2$-GaMe} is to estimate the class of cluster galaxy members to get an unbiased estimation of physical properties, such as dynamical mass measurements, to constrain cosmology and astrophysics using upcoming galaxy surveys. The main results are summarized below: 
\begin{itemize}
\item We showed that \textsc{C$^2$-GaMe}'s probabilistic classification scheme based on RF is calibrated and has a better statistical and physical performance than Logistic and KNN classifiers. 

\item We demonstrate the probabilistic classification scheme reproduces the radial and velocity distribution of galaxies with $<3\%$ error (see \Cref{fig:r0_dists_2d,fig:v_dist_match_infall,fig:p_dists_2d}). The deterministic classifiers, however, estimate galaxies' radial and velocity distribution with $\sim10\%$ error. We recommend using the probabilistic model for accurate prediction of the spatial and velocity distribution of galaxies. 

\item We showed that the probabilistic output of \textsc{C$^2$-GaMe} gives us an unbiased estimation of dispersion attainable from projected space information of orbiting and infalling galaxies in the presence of interlopers. \textsc{C$^2$-GaMe} predicts the velocity dispersion squared of cluster galaxies with $\lesssim1\%$ accuracy, out-performing the physical model presented in \citet{Aung22} with $\lesssim4\%$ accuracy (see \Cref{fig:vd_ratio_mass_dependence}).
\item We showed that adding other observables, such as the mass ratio of the galaxy stellar mass to the halo of the cluster and specific star formation rate (sSFR) as an additional input feature, increases the accuracy to $80\%$ from $77\%$ with just projected radius and line-of-sight velocity. 
\item We finally tested the robustness of \textsc{C$^2$-GaMe} trained on MDPL2+UM galaxy catalogs to the simulated galaxies extracted from Illustris-TNG hydrodynamical simulations. We found that the probabilistic classification reproduces the probability distribution function with $<5\%$ difference outside $0.5\r200m$ and $<10\%$ inside when using position and velocity information only (see \Cref{fig:validation}). However, adding sSFR as an additional feature degrades the performance as the different simulations do not converge on the same distribution. 
\end{itemize}

The implementation of the \textsc{C$^2$-GaMe} algorithms and the trained models is publicly
available in a GitHub repository \href{https://github.com/DannyFarid/C2-GaMe-Classification-of-Cluster-Galaxy-Membership}{link}. This python package allows the user to perform cluster galaxy membership classification seamlessly. 

\section*{Acknowledgement}
We acknowledge Dhayaa Anbajagane for sharing the preprocessed TNG galaxy catalogs used in this work. We also thank Benedikt Diemer and Johnny Esteves for their helpful comments on the draft. We thank the anonymous reviewers for their helpful suggestions and comments on the manuscript. We also acknowledge the use of Python 3 in a Jupyter Notebook \citep{jupyter}, scikit-learn \citep{scikit-learn} for the machine learning models, and matplotlib \citep{matplotlib} for plotting, and numpy \citep{numpy} and pandas \citep{pandas} for management and manipulation of the data. This work was supported by Yale University and the facilities and staff of the Yale Center for Research Computing. DN also acknowledges support by NSF (AST-2206055) and NASA (80NSSC22K0821 \& TM3-24007X) grants. 

\section*{Data Availability}
The MDPL2 halo catalogs and the mock UM galaxy catalogs are publicly available at \url{https://www.peterbehroozi.com/data.html}. The TNG halo catalog is available at \url{https://www.tng-project.org/data/}. The models are available at \textsc{C$^2$-GaMe} GitHub repository\footnote{\url{https://github.com/DannyFarid/C2-GaMe-Classification-of-Cluster-Galaxy-Membership}}.

\appendix

\section{Classification in 3D Phase Space}
\label{sec:3D}
\begin{figure}[ht]
	\includegraphics[width=0.5\textwidth]{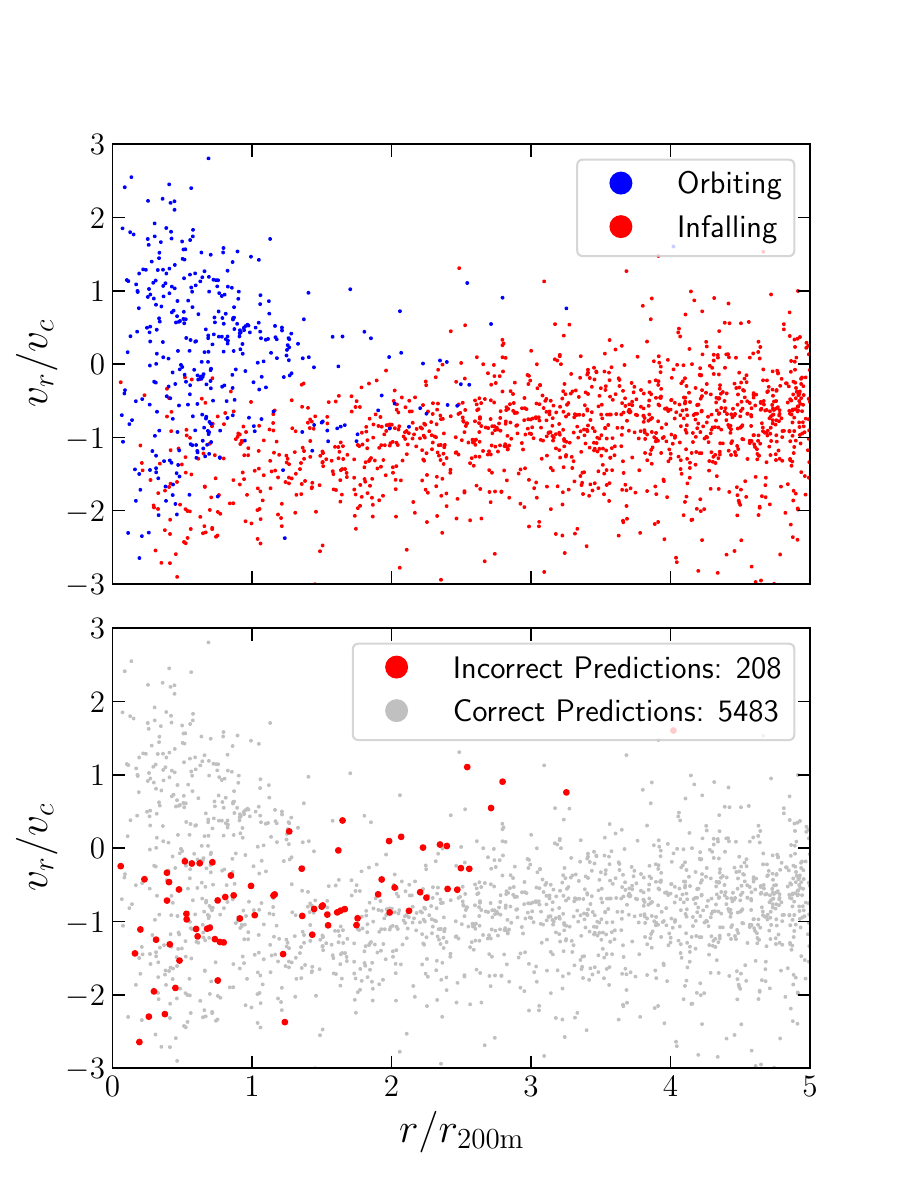}
	\caption{ \textit{Top panel}: distribution of the galaxies in radial velocity $v_r$ and 3D radius $r$, visually showing how the galaxies are arranged in the dark matter halo. \textit{Bottom panel}: distribution of ML predictions using RF showing the distribution of correct and incorrect predictions when applying deterministic classification. Even though we do not input $v_r$ and $r$ as a feature, but indirectly use the 3 positions and 3 velocities, ML correctly predicts galaxies with positive radial velocity as orbiting, and the incorrect predictions lie where there is no clear separation in phase space similar to 2D result. The figure is downsampled from journal version.}
    \label{fig:vr_r_3d}
\end{figure}

We also use the three normalized positional dimensions $x,y,z/\r200m$ and three normalized velocity dimensions $v_x,v_y,v_z/v_c$ as the input features to classify the galaxies/halos inside the turnaround radius as either orbiting or infalling.  

The deterministic classification with the RF model yields 96.4\% accuracy in its classification task. \Cref{fig:vr_r_3d} shows where incorrect classifications occur in the phase space diagram. We can see a clear separation of orbiting and infalling galaxies where the orbiting galaxies are distributed preferentially in the inner region of $0<r/\r200m<2$, while the infalling galaxies are distributed preferentially at larger radii to the maximum limit in our sample of $5\r200m$. However, there is a notable region of overlap where the two populations are mixed, where the radial velocity is negative inside $2r_{\rm 200m}$. We expect this region to be prone to incorrect predictions, and the top panel of \Cref{fig:vr_r_3d} confirms this as most incorrect predictions occur between $0 \leq r/r_{200m} \leq 2$ and $-3 \leq v_r/v_c \leq 0$. 

\begin{figure}
	\includegraphics[width=0.5\textwidth]{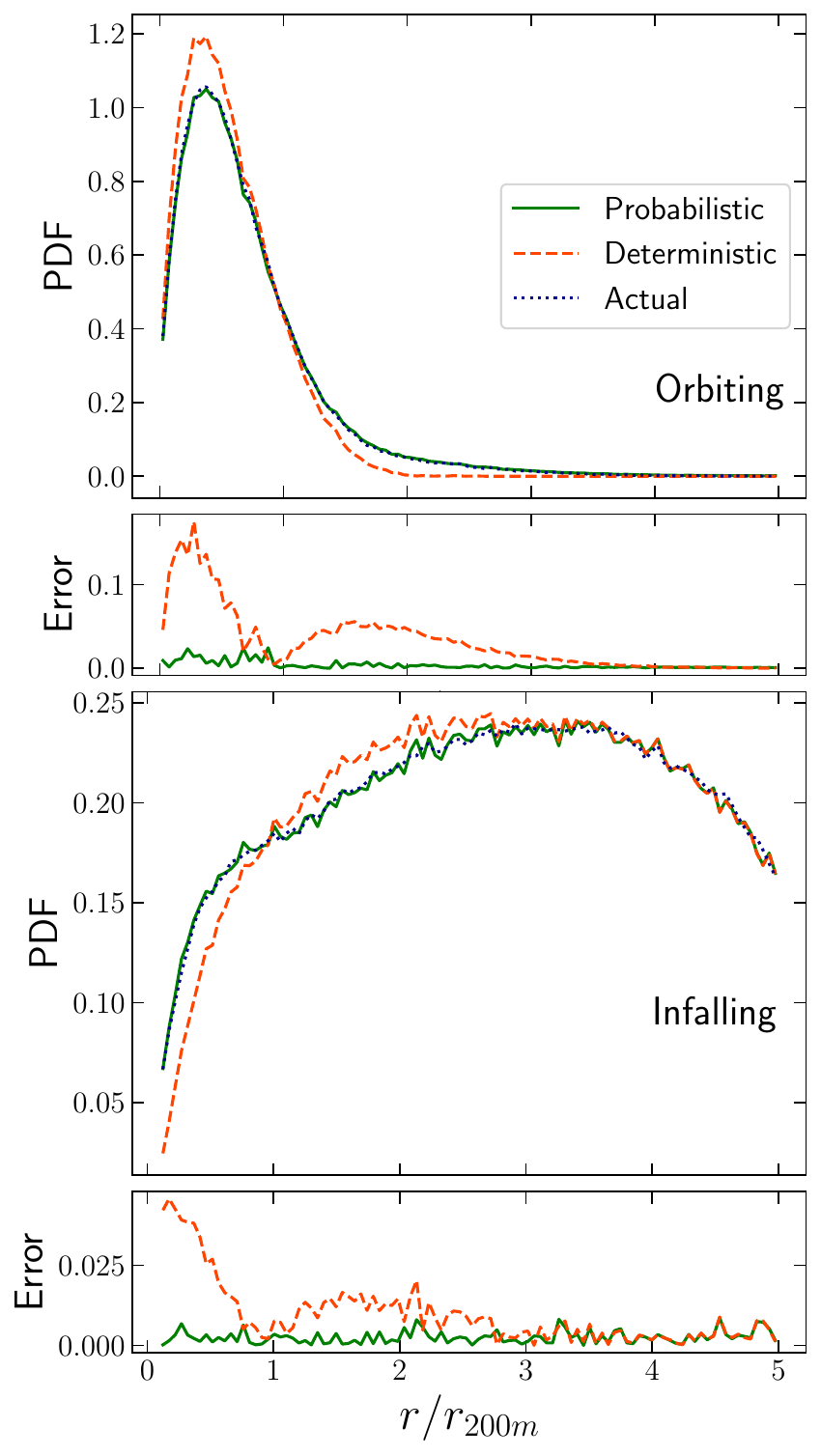}
    \caption{
    Comparison of the probability density function as a function of 3D radius $r$ between deterministic classification prediction (orange), probabilistic prediction (green), and actual distribution (blue) for the orbiting (\textit{left panel}) and infalling (\textit{right panel}) galaxy population using 3D data, where the \textit{top panel} shows the probability density function, while the \textit{bottom panel} shows the absolute difference between actual and predicted PDF. Similar to the classification in projected phase space in \Cref{fig:r0_dists_2d}, the probabilistic distribution (green) reproduces the actual distribution with $<1\%$ error. The deterministic classification method is similar to the method used in \citet{delosRios2021}.} 
    \label{fig:r_0_dists}
\end{figure}

We also compare the deterministic and probabilistic classification schemes by comparing the actual and classified probability density function (PDF) of orbiting and infalling halos. Under these two schemes, we look at the radial ($r/r_{200m}$) distributions for each classification of the galaxy (infalling or orbiting). Comparing these distributions to the actual radial distribution for either classification shows us the overall performance of the RF model and the relative performance between deterministic classification and probabilistic. \Cref{fig:r_0_dists} shows the differences between the $r/r_{200m}$ distributions of the deterministic classification scheme, the probabilistic scheme, and the actual distribution. The predicted distribution from the probabilistic scheme is much closer to the actual distribution, as confirmed in the figure by the relatively lower error. We see a similar trend for the distribution of $r/r_{200m}$ for the infalling population. Since the phase space information is insufficient to split the orbiting and infalling galaxies in this region, we do not expect deterministic classification to match the actual distribution perfectly. However, probabilistic classification addresses this problem by attributing appropriate probability for each galaxy in the phase space so that the correct number of galaxies is counted on average. 

\bibliographystyle{elsarticle-harv}
\bibliography{bib}

\end{document}